\documentclass[a4paper,aps,prd,10pt,preprintnumbers,showpacs,twocolumn,superscriptaddress,nofootinbib,amsmath,amssymb,floatfix]{revtex4-1}\def\JCAPstyle#1{}

\usepackage{graphicx}
\usepackage[utf8]{inputenc}
\usepackage[T1]{fontenc}
\usepackage{cmap}

\DeclareMathAlphabet{\pazocal}{OMS}{zplm}{m}{n}

\begin{document}

\preprint{APS/-QNM}

\title{Quantum corrections to the quasinormal modes of the Schwarzschild black hole}
\author{Hao Chen}
\email{haochen1249@yeah.net}
\affiliation{School of Physics and Electronic Science, Zunyi Normal University, Zunyi 563006, China}
\affiliation{College of Physics, Guizhou University, Guiyang, 550025, China}
\author{Hassan Hassanabadi}
\email{h.hasanabadi@shahroodut.ac.ir}
\affiliation{Faculty of Physics, Shahrood University of Technology, Shahrood, Iran}
\affiliation{Department of Physics, University of Hradec Kr$\acute{a}$lov$\acute{e}$, Rokitansk$\acute{e}$ho 62, 500 03 Hradec Kr$\acute{a}$lov$\acute{e}$, Czech Republic}
\author{Bekir Can L\"{u}tf\"{u}o\u glu}
\email{bclutfuoglu@akdeniz.edu.tr}
\affiliation{Department of Physics, University of Hradec Kr$\acute{a}$lov$\acute{e}$, Rokitansk$\acute{e}$ho 62, 500 03 Hradec Kr$\acute{a}$lov$\acute{e}$, Czech Republic}
\affiliation{Department of Physics, Akdeniz University, Campus, 07058 Antalya, Turkey}
\author{Zheng-Wen Long}
\email{ zwlong@gzu.edu.cn}
\affiliation{College of Physics, Guizhou University, Guiyang, 550025, China}
\date{\today }

\begin{abstract}
Based on the minimum measurable momentum concepts associated with the quantum gravity effects acting on the large-scale dynamics of the universe, we study the quantum effect of the EUP on the Hawking evaporation of the black hole. The results show the quantum corrections may shorten the lifetime of the massive black hole. To verify the new EUP on the black hole stability, the scalar field and electromagnetic field are derived and the time evolution of the black hole is analyzed in terms of the time domain integration method, the quantum effect alters the oscillation and decay time of black hole. Furthermore, we use the WKB numerical approximation method to calculate the quasinormal mode frequencies and analyze the influence of the EUP parameter $\alpha$ on the scattering problem. This shows that the EUP significantly increases the area of the total absorption cross-section of the black hole.
\end{abstract}

\maketitle

\section{Introduction}
In 1915, Einstein proposed the general theory of relativity where gravity is assumed to be a geometric property of space and time \cite{Ein1}. Since then, this theory is regarded as the best description of gravity. In general relativity, one set of solutions to Einstein's field equations indicates the existence of black holes.  The black hole is one of the mysterious facts of the universe, producing an effective area that even light cannot escape. It is believed that black hole is formed in the final stage of the gravitational collapse of stars \cite{bh1,bh2}. Black holes are considered a laboratory for investigating gravity because they can emit gravitational radiation associated with their oscillations. Such radiation is assumed to be very important since it would carry internal information about the black hole's characteristics, such as its mass and charge \cite{haw1,haw2}. Nowadays, quantum phenomena are assumed to have a vital role in the black hole radiation. The perturbation theory of black holes is the interaction between the external field and black hole, which mainly includes the scalar quasinormal (Q-N) modes \cite{scalar1}, and the electromagnetic quasinormal Q-N modes \cite{electr2} and the gravitational quasinormal Q-N modes \cite{electrr4}. Recently, the LIGO-VIRGO cooperation proposed a method to directly detect the gravitational waves from mergers of black holes and neutron stars and find the ringdown in the gravitational wave \cite{BF1,BF2}, which sparks intense interest in Q-N modes. The authors in Ref. \cite{jx1,jx2} considered the Q-N modes and greybody bounds to detect the black holes. The authors in Ref. \cite{jx3} used to GUP quantum effect to investigate the ringing of the black hole.

Vishveshwara described a type of spacetime perturbation that depends on particular conditions, near the black hole event \cite{canek1}. According to his paper for a set of complex-valued frequencies, soon  afterwards these frequencies is also known as the Q-N frequencies \cite{canek2} and have been extensively examined \cite{kk2,kk4,kk5,dong1}. In those studies, it is shown that the real part of a  Q-N frequency determines the oscillation frequency while its imaginary part  means the damping rate \cite{canek3}. The Wentzel-Kramers-Brillouin (WKB) approximation is one of the mathematical tools used for analyzing Q-N modes. Schutz and Will were the pioneers and they employed this approach up to the first order \cite{kk6}. Then, the method is extended up to the second order in \cite{kk7,kk8} and lately up to the sixth order \cite{kk9}. As an alternative method to WKB, Leaver used numerical methods to calculate the gravitational Q-N modes of stationary and rotating black holes \cite{kk10,kk11}. A brief summary of further and recent methods that produce accurate results in the determination of Q-N modes can be found in \cite{kk13,kl1,long1,long2}.

In general relativity, the existence of singularities corresponding to the solutions of Einstein's field equations associated with black holes will promote the breakdowns of general relativity theory. To eliminate the existing singularity problems, a lot of ideas have been proposed. According to one of them, a quantum deformation can be used for the black hole  to obtain the solution of the regular black hole which was given by Bardeen in 1968 \cite{kk14}.  To this end,  the authors in Ref. \cite{kk15} showed that the singularity at $r=0$ is shifted to a finite radius value \cite{kk15}. The authors in Ref. \cite{kk17} explored the Hawking radiation of quantum black holes Furthermore, some authors studied the thermodynamics  properties and weak deflection angle of the  different black holes according to the modified uncertainty principles in \cite{kk18,kk19,kk20,kk21,kk22,kkk1,kk23, chen1, ay2}.
Based on the noncommutative geometry theory \cite{kk24,kk25}, the authors in Ref. \cite{kk26,kk27,kk28,kk29} applied it to investigate the thermodynamic properties of black holes. Besides, Anacleto et al. further explored the scattering problem for the noncommutative corrected black hole\cite{kk30}.

On the other hand, we observe that quantum deformation usage is not limited to these studies, papers that examine Q-N modes in the presence of the quantum deformation becomes popular recently. For example, last year Konoplya studied the Q-N modes of a quantum-modified black hole \cite{kk16}. Based on the existence of the minimal length, Kempf proposed the generalized uncertainty principle (GUP) in terms of the Heisenberg Algebra \cite{GUP1}, Anacleto et al. also studied the Q-N modes and shadow of the black hole and calculated the Q-N frequencies of the black hole up to the sixth-order WKB method \cite{Anacleto}. Similar work can be found in Ref. \cite{kk31,kk33,kk34,ay1}. In this work, our motivation is to investigate the new EUP-corrected evaporation rate to explore the stability of black holes. To this end, we further study the Q-N frequencies of the black hole to verify its stability of the black hole. In addition, we also analyze the impact of the new EUP on the black hole scattering problem to enrich the content. The paper is organized as follows: In Sec. 2, we investigate the evaporation rate of the new EUP corrected Schwarzschild black hole. In Sec. 3,  we analyze the effect of quantum correction on the dynamical evolutions of the black hole for the scalar field and electromagnetic field perturbation and calculate the Q-N frequency by using the  3rd order WKB approach. In Sec. 4, the influence of the EUP parameter $\alpha$ on the absorption cross section is discussed. The final comments are in Sec.5.
\section{Quantum effects of the Hawking evaporation}\label{sec: ER}
In this section, we start by considering the following $1$ dimensional deformed Heisenberg algebra (EUP) that leads to measuring a minimum momentum value \cite{hh}:
\begin{equation}
[X, P] \geqslant \frac{i \hbar}{1-\alpha|X|},
\end{equation}
here, $\alpha$ is the deformation parameter which takes value in the range of $0 \leq \alpha\leq 1$. Following this deformation,
we find
\begin{equation}
\begin{aligned}
\Delta X \Delta P & \geqslant \frac{\hbar}{2}\left\langle\frac{1}{1-\alpha|X|}\right\rangle \\
& \geqslant \frac{\hbar}{2}\left[1+\alpha\langle|X|\rangle+\alpha^{2} X^{2}+q^{3}|X| X^{2}+\cdots\right]\\ & \geqslant  \frac{\hbar}{2}[-\alpha(\Delta X)+\alpha(\Delta X)+1+\alpha\langle|X|\rangle+\cdots]\\
& \geqslant  \frac{\hbar}{2}\left[-\alpha(\Delta X)+\frac{1}{1-\alpha(\Delta X)}\right].
\end{aligned}
\end{equation}
Then, we express the momentum uncertainty in terms of position uncertainty. We arrive at
\begin{equation}\label{eup1}
\Delta P \geq \frac{\hbar}{2 \Delta X}\left[-\alpha(\Delta X)+\frac{1}{1-\alpha(\Delta X)}\right].
\end{equation}
It can be found that the new EUP gives the minimal momentum $\Delta P \geqslant \frac{3}{2} \alpha \hbar$. If we take the uncertainty parameter $\alpha=0$, the equation \eqref{eup1} recovers the Heisenberg uncertainty principle $\Delta X \Delta P \geq \frac{\hbar}{2}$. The Hawking temperature corresponding to any massless quantum particle near the Schwarzschild black hole event horizon is given by (the units $\hbar=c=G=1$)
\begin{equation}
T=\frac{\Delta P}{\kappa},
\end{equation}
here $\kappa$ is the Boltzmann constant. Based on this, we can find that the lower bound of the Hawking temperature reads
\begin{equation}
T\geq T_{min}=\frac{3}{2}\frac{\alpha}{\kappa}.
\end{equation}
Based on the properties of the uncorrected horizon radius $r_{h}$ and uncertain position $\Delta X$ in Ref.\cite{kkqq1,x1,x2}
\begin{equation}
\Delta X\simeq 2\pi r_{h }=4\pi M.
\end{equation}
In this case, we can calculate the general total of the mass-temperature
\begin{equation}
M=\frac{1}{8 \pi \alpha}\left[1-\left(1-\frac{4 \alpha}{\alpha+2 \kappa T}\right)^{\frac{1}{2}}\right].
\end{equation}
For a small EUP parameter value $\alpha$, the new EUP-corrected Hawking temperature can be expressed as
\begin{equation}
T_{EUP}=\frac{4\pi\alpha M-1-16\pi^{2} \alpha^{2} M^{2} }{8 \pi \kappa M\left(4 \pi \alpha M -1\right)}.
\end{equation}
If we ignore the effect of quantum effects on the black hole $(\alpha=0)$, the Hawking temperature is given by
\begin{equation}
T=\frac{1}{8 \pi M \kappa},
\end{equation}
this result is completely consistent with the Ref \cite{kk19}. In addition, the area of horizon $A$ can be given by
\begin{equation}
A=4 \pi r^{2}_{h}=16\pi M^{2}.
\end{equation}
Based on this ,we consider the Stefan-Boltzmann law \cite{gs1,gs2}
\begin{equation}
\frac{d E}{d t}=\sigma A T^{4},
\end{equation}
here $\sigma = \pi^{2}/60$ is the Stefan-Boltzmann constant. Furthermore, we take the evaporated mass of the black hole is completely converted into energy:
\begin{equation}
d E=d M.
\end{equation}
In this case, the evaporation rate of the new EUP corrected Schwarzschild black hole can be expressed as
\begin{equation}
\begin{aligned}
\frac{d M}{d t} & \simeq \frac{4 \pi^{3} M^{2}}{15} T_{E U P}^{4} \\
& \simeq \frac{4 \pi^{3} M^{2}}{15}\left(\frac{4 \pi \alpha M-1-16 \pi^{2} \alpha^{2} M^{2}}{8 \pi \kappa M(4 \pi \alpha M-1)}\right)^{4}.
\end{aligned}
\end{equation}
\begin{figure}
\includegraphics[width=0.40\textwidth]{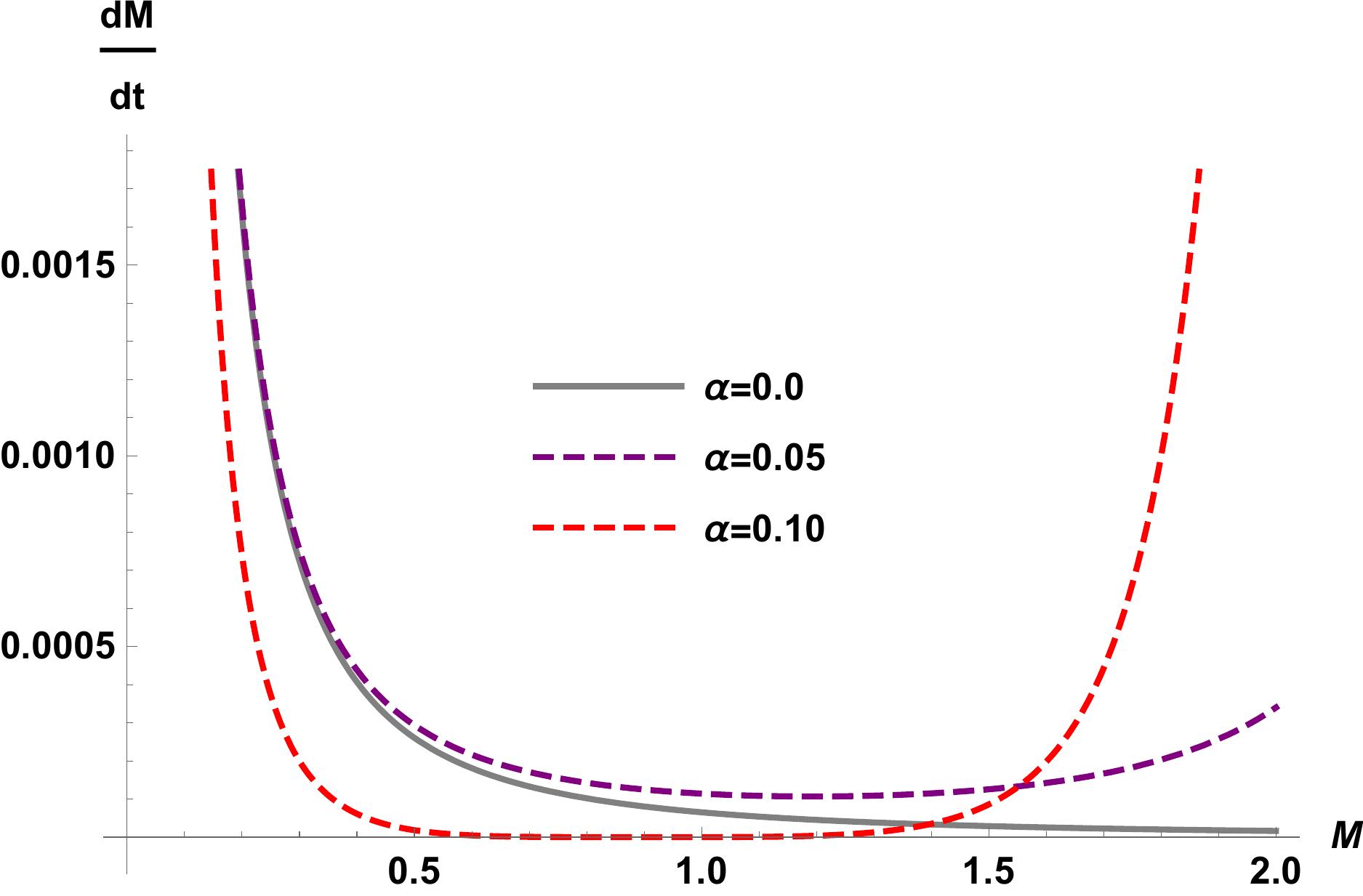}
\caption{The evaporation rate of the new EUP-corrected black hole as the functions on M for the different EUP parameters $\alpha$: $(\kappa=1)$.}
\label{fig:1}
\end{figure}
From the behaviour shown in FIG. 1, we can see that the uncorrected Schwarzschild black hole has a maximum evaporation rate at a lower mass. As the mass increases, the black hole tends to be larger time to decay. However, the new EUP parameter $\alpha$ has a non-negligible effect on the evaporation rate in the massive black holes, and it is concluded that quantum correction associated with the new EUP may drastically shorten the lifespan of the Schwarzschild black holes.
\section{The Q-N modes of the new EUP corrected Schwarzschild black hole.}\label{sec:QNMs}
In this section, we will give the line element of the EUP corrected black hole so that the  Q-N modes can be studied. Therefore, we consider a massless particle and assume $\Delta P \sim P \sim E$, and find the following bound:
\begin{equation}
E \Delta X \geq \frac{1}{2},
\end{equation}
which leads  to the new EUP in equation \eqref{eup1} to be written in the form of
\begin{equation}
\begin{aligned}
\varepsilon & \geq E\left[-\alpha(\Delta X)+\frac{1}{1-\alpha(\Delta X)}\right] \\
& \geq E\left[-\alpha r_{h}+\frac{1}{1-\alpha r_{h}}\right].
\end{aligned}
\end{equation}
Here,  $\varepsilon$ denotes the black hole corrected energy due to the EUP. We follow the arguments given in \cite{x1,x2} and consider the position uncertainty nearly equal to the horizon radius, $\Delta X\sim 2r_{h}$. Then, we rewrite equation \eqref{eup1} by considering $E \sim M$, $\varepsilon=M_{EUP}$, and $r_{h}=2M$. We find
\begin{equation}\label{Meupcan}
\begin{aligned}
M_{E U P} & \geq M\left[-\alpha r_{h}+\frac{1}{1-\alpha r_{h}}\right] \\
&=M\left[-2 \alpha M+\frac{1}{1-2 \alpha M}\right],
\end{aligned}
\end{equation}
where $M_{E U P}$ is the EUP-corrected mass function of the black hole. Then, we derive the following relation  for the event horizon out of the equation \eqref{Meupcan}
\begin{equation}
r_{h E U P}=2 M_{E U P}=M\left[-4 \alpha M+\frac{2}{1-2 \alpha M}\right].
\end{equation}
Based on these findings, we modify the Schwarzschild black hole metric, the EUP-corrected line element reads
\begin{equation}\label{Meupmetric}
\begin{aligned}
d s^{2} &=\left(1-\frac{2 M_{\mathrm{EUP}}}{r}\right) d t^{2}-\left(1-\frac{2 M_{\mathrm{EUP}}}{r}\right)^{-1} d r^{2} \\
&-r^{2}\left(d \theta^{2}+\sin ^{2} \theta d \phi^{2}\right).
\end{aligned}
\end{equation}
Next, we will analyze the influence of the scalar field and electromagnetic disturbance on the EUP-corrected Schwarzschild black hole.\\ For a massless scalar field in curved space-time can be described by the Klein-Gordon equation \cite{scalar1}:
\begin{equation}\label{zmywan1}
\frac{1}{\sqrt{-g}} \partial_{\mu}\left(\sqrt{-g} g^{\mu \nu} \partial_{\nu} \Psi\right)=0.
\end{equation}
For the electromagnetic field equation\cite{electr2}
\begin{equation}\label{mjl2}
\frac{1}{\sqrt{-g}} \partial \nu\left(F_{\rho \sigma} g^{\rho \mu} g^{\sigma \nu} \sqrt{-g}\right)=0,
\end{equation}
here$F_{\rho \sigma}=\partial \rho A^{\sigma}-\partial \sigma A^{\rho}$, $A_{\nu}$ is the vector electromagnetic field potential. Due to the symmetry of the corrected black hole \eqref{Meupmetric}, we use the separation variable method and suppose the wave function satisfies the following form:
\begin{equation}
\Psi_{\omega l m}(\mathbf{r}, t)=\frac{R_{\omega l}(r)}{r} Y_{l m}(\theta, \phi) e^{-i \omega t},
\end{equation}
where $Y_{l m}(\theta, \phi)$ denotes the spherical harmonics and $\omega$ indicates the quasinormal mode frequencies. By defining the tortoise coordinate
\begin{equation}
d r_{*}=\frac{1}{1-\frac{2 M_{\text {EUP }}}{r}} d r,
\end{equation}
we can derive the general Schr\"odinger equation
\begin{equation}\label{Mastereq}
\frac{d^{2} R_{\omega l}\left(r_{*}\right)}{d r^{2}}+\left[\omega^{2}-V(r)\right] R_{\omega l}\left(r_{*}\right)=0,
\end{equation}
the effective potential of the scalar field $V_{e f f}(r)$ and the electromagnetic field $V_{em}(r)$ read
\begin{equation}\label{Veffpot}
V_{e f f}(r)=\left(1-\frac{M_{E U P}}{r}\right)\left(\frac{l(l+1)}{r^{2}}+\frac{1}{r} \frac{d\left(1-\frac{M_{E U P}}{r}\right)}{d r}\right).
\end{equation}
and
\begin{equation}\label{Veffpot1}
V_{em}(r)=\left(1-\frac{M_{E U P}}{r}\right)\left(\frac{l(l+1)}{r^{2}}\right).
\end{equation}
Here, $l$ indicates the multipole number of the spherical harmonics. Next, we will use the time domain integration method to investigate the scalar field and the electromagnetic  field perturbation of the black hole \cite{electr3,electr4}, the wave function and the effective potential of the black hole are constrained by
\begin{equation}
\psi\left(r_{*}, t\right)=\psi\left(i \Delta r_{*}, j \Delta t\right)=\psi_{i, j},
\end{equation}
\begin{equation}
 V\left(r\left(r_{*}\right)\right)=V\left(r_{*}, t\right)=V_{i, j},
\end{equation}
In this case, the massless scalar field in equation \eqref{zmywan1} can be re-expressed as
\begin{equation}
\begin{aligned}
&\frac{\psi_{i+1, j}-2 \psi_{i, j}+\psi_{i-1, j}}{\Delta r_{*}^{2}}-\frac{\psi_{i, j+1}-2 \psi_{i, j}+\psi_{i, j-1}}{\Delta t^{2}} \\
&-V_{i} \psi_{i, j}=0.
\end{aligned}
\end{equation}
Now, we consider initial conditions \cite{electr5}
\begin{equation}
\psi\left(r_{*}, t\right)=\exp \left[-\frac{\left(r_{*}-k_{1}\right)^{2}}{2 \sigma^{2}}\right], \quad \left.\psi\left(r_{*},  t\right)\right|_{t<0}=0,
\end{equation}
where $k_{1}$ and $\sigma$ are the median and width of the initial wave- packet. Based on this, time evolution of the scalar field reads
\begin{equation}
\begin{aligned}
\psi_{i, j+1}&=-\psi_{i, j-1}+\left(\frac{\Delta t}{\Delta r_{*}}\right)^{2}\left(\psi_{i+1, j+\psi_{i-1, j}}\right)\\
&+\left(2-2\left(\frac{\Delta t}{\Delta r_{*}}\right)^{2}-V_{i} \Delta t^{2}\right) \psi_{i, j},
\end{aligned}
\end{equation}
To satisfy the Von Neumann stability condition, the relation is maintained in the numerical process: $(\frac{\Delta t}{\Delta r_{*}}<1)$. In the same way, we can use the same method to study the EUP corrected black hole perturbed by the electromagnetic field.
\begin{figure*}
\resizebox{\linewidth}{!}{\includegraphics{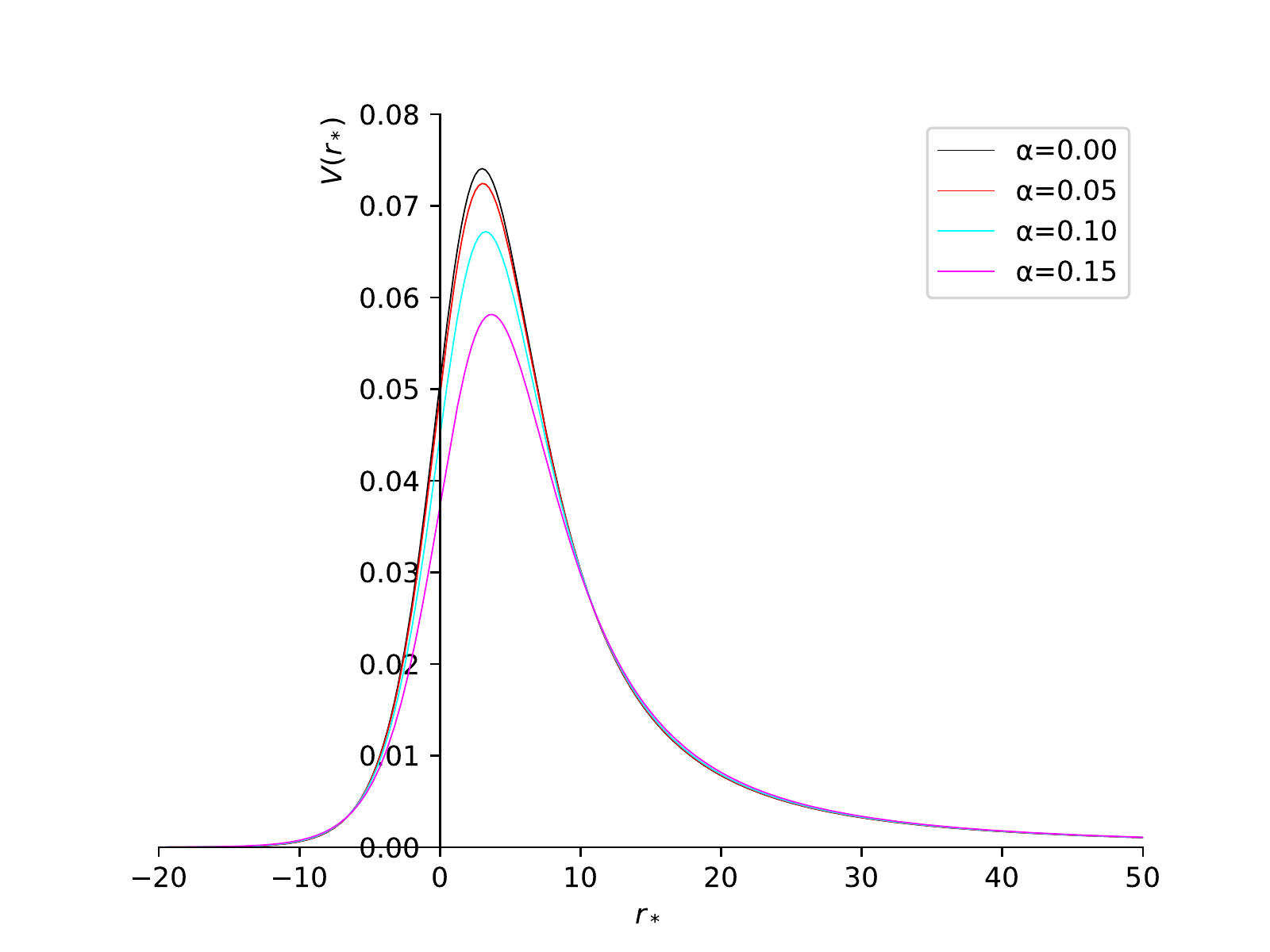},\includegraphics{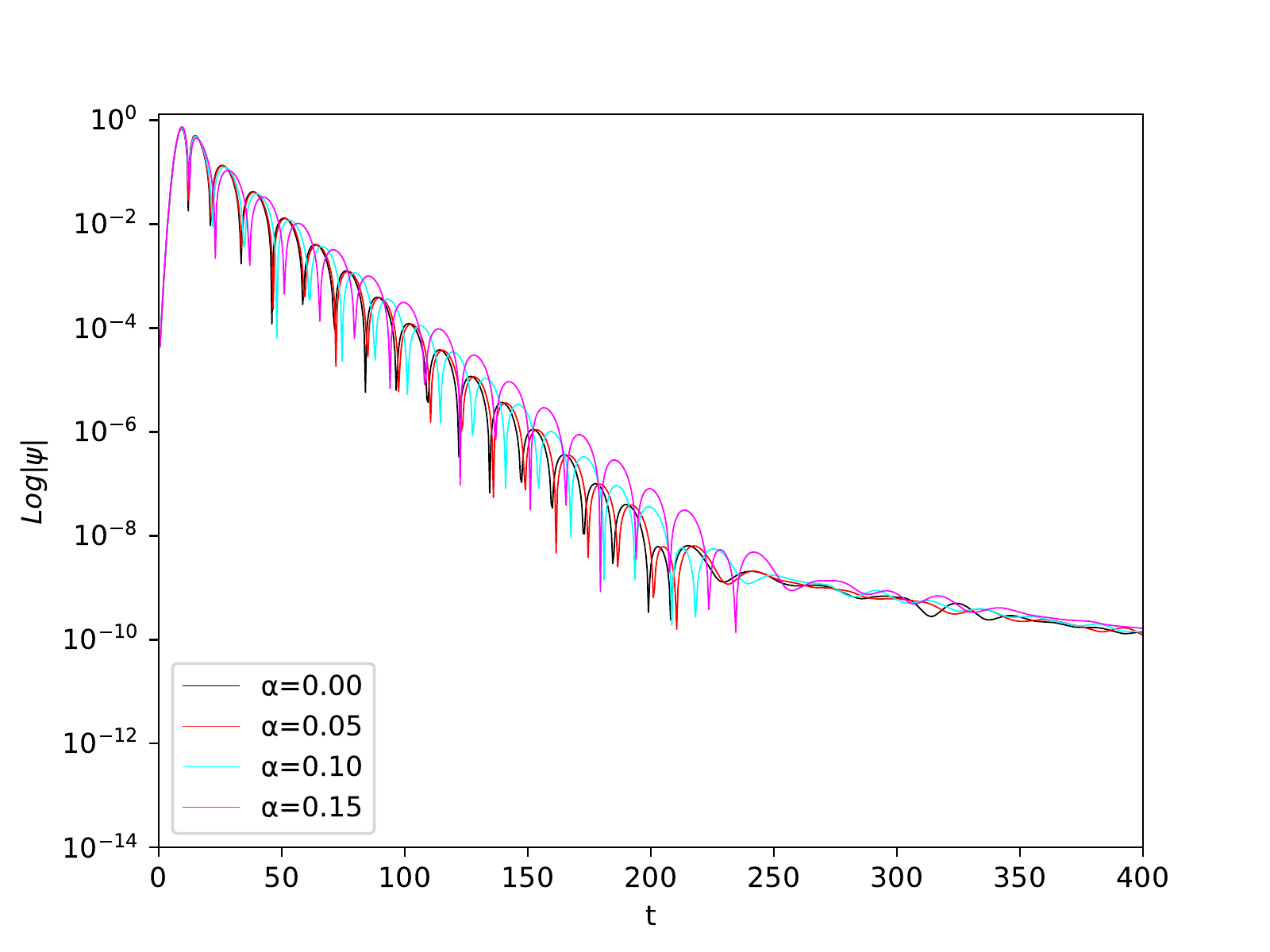}}
\caption{Left panel: The effective potential of the scalar field perturbation in the EUP corrected Schwarzschild black hole. Right panel: The time-domain profiles of the  scalar field perturbation in the EUP corrected Schwarzschild black hole.}\label{fig:t1}
\end{figure*}
\begin{figure*}
\resizebox{\linewidth}{!}{\includegraphics{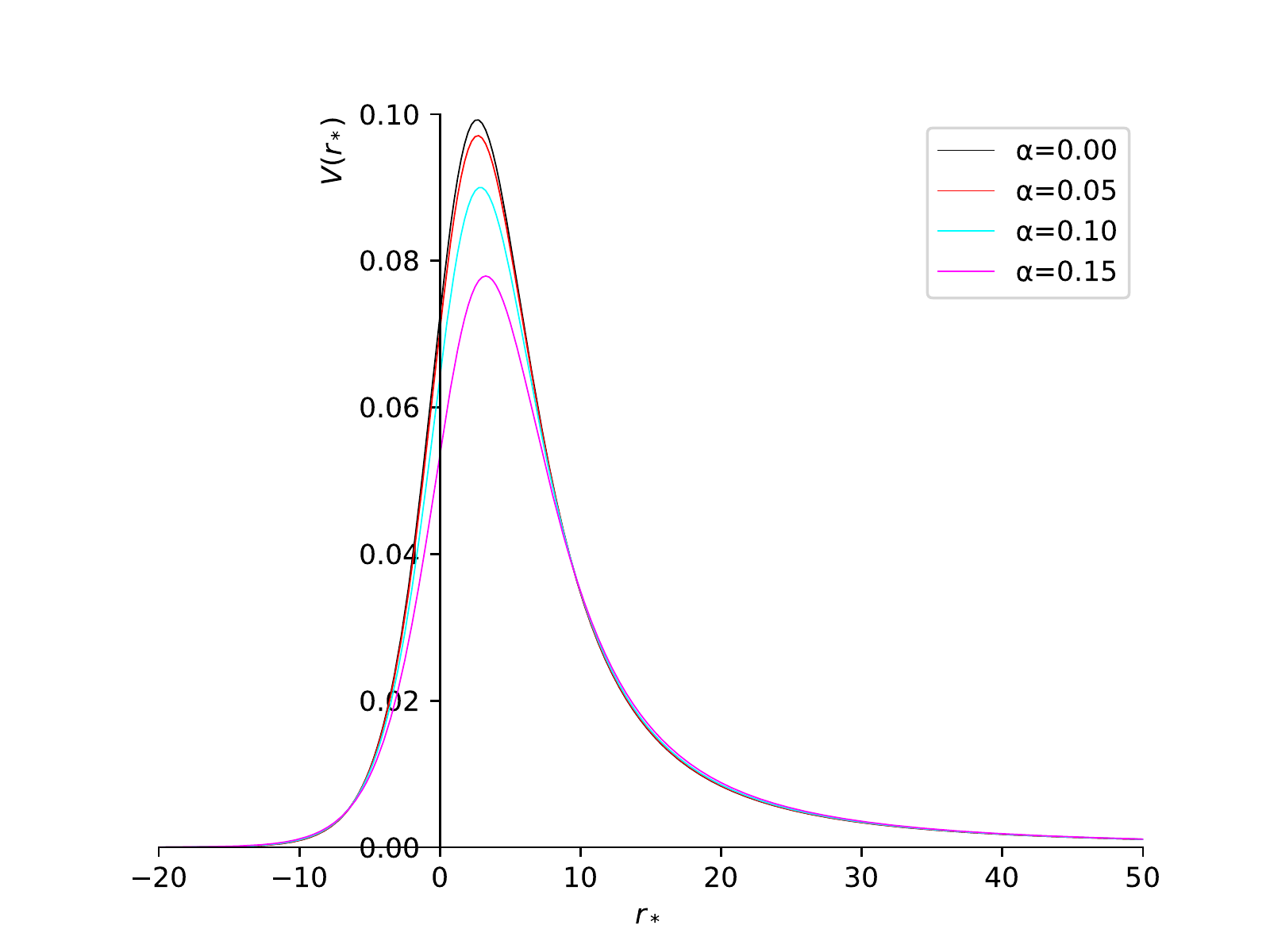},\includegraphics{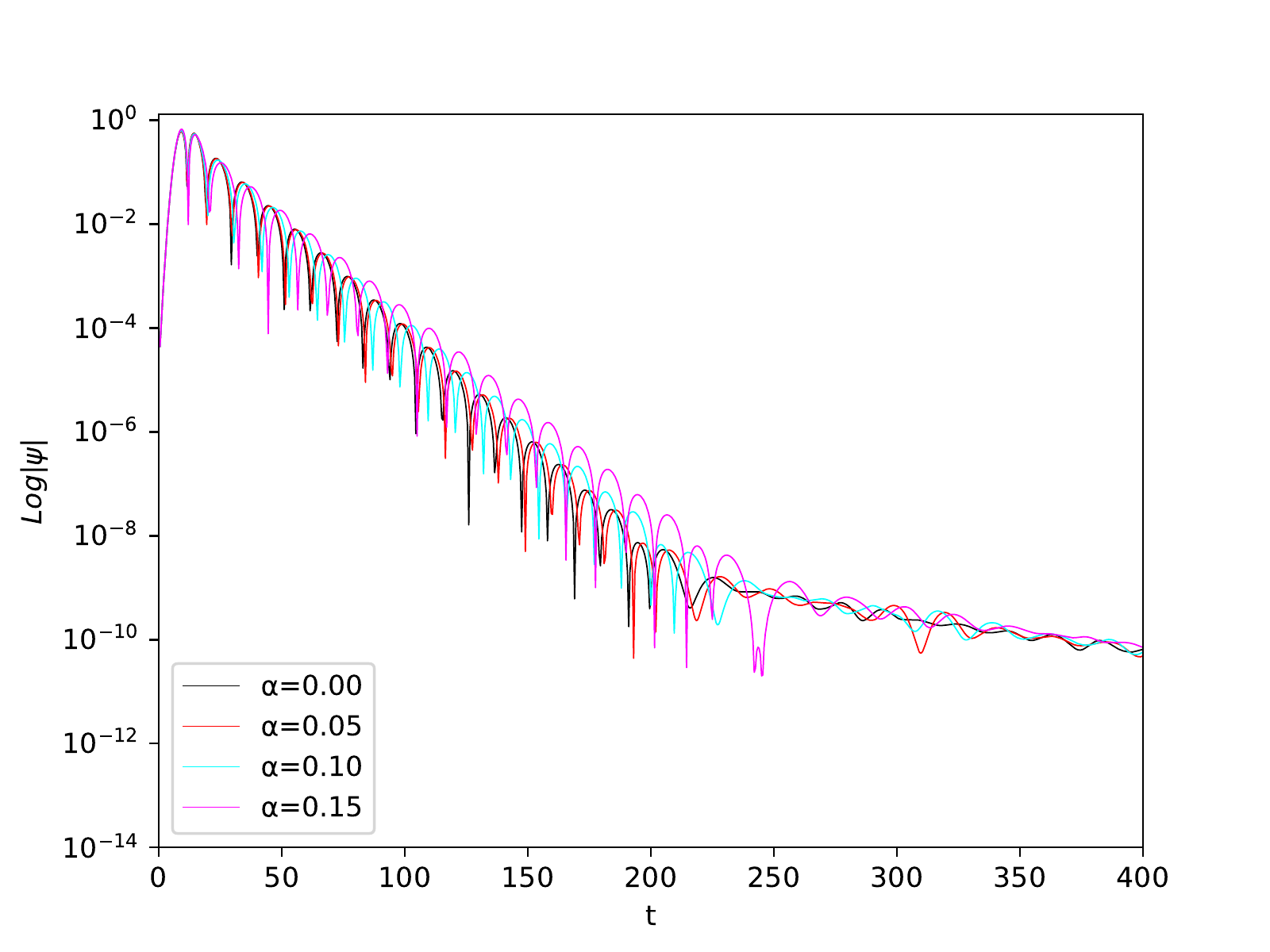}}
\caption{Left panel: The effective potential of the electromagnetic field perturbation in the EUP corrected Schwarzschild black hole. Right panel: The time-domain profiles of the electromagnetic field perturbation in the EUP corrected Schwarzschild black hole.}\label{fig:t1}
\end{figure*}
From the behaviour shown in FIG.2-3, we give the effective potential and the time-domain profiles of the scalar field and electromagnetic fields, where $M=0.5,l=1$. On the one hand, we can observe that the effective potential decreases as the EUP parameter $\alpha$ increases, when the tortoise coordinate $r_{*}$ tend to infinity, the effective potential approaches zero and the black hole returns to an equilibrium state, the different field perturbations just change the barrier peak height. On the other hand, we find as the EUP parameter increases, the oscillation and decay time of the black hole become slower, the results show that the  Q-N modes of the Schwarzschild black hole is not determined by the perturbation  model.\\

Next, we will investigate the Q-N frequency with which the scalar field perturbs the black hole, the Q-N frequency can be computed by solving the wave equation \eqref{Mastereq}, requiring pure outgoing waves at infinity and at the event horizon:
\begin{equation}
\Psi \propto e^{-i \omega t \pm i \omega r_{*}}, \quad r_{*} \rightarrow \pm \infty.
\end{equation}
\begin{table*}
\begin{tabular}{|c|c|c|c|c||c||c||c||c|}
\hline

\hline
$l=1$\\
\hline
 $\alpha$ &$\omega_{0}$ & $\omega_{1}$ &  $\omega_{2}$  \\
\hline
\hline
$0.01$   & 0.288115 - 0.096992 i & 0.259511 - 0.304266  i & 0.221240 - 0.521391 i \\
$0.02$   &  0.284940 - 0.095923 i  & 0.256651 - 0.300913 i & 0.218802 - 0.515645 i \\
$0.03$   & 0.281588 - 0.094795  i  & 0.253632 - 0.297372 i & 0.216228 - 0.509578 i  \\
$0.04 $   & 0.278058 - 0.093606 i  & 0.250452 - 0.293644 i & 0.213517 - 0.503189 i \\
$0.05$  & 0.274348 - 0.092357 i  & 0.247111 - 0.289727 i & 0.210669 - 0.496477 i  \\
$0.06$    & 0.270461 - 0.091049 i  & 0.243609 - 0.285621  i & 0.207683 - 0.489442 i  \\
$0.07 $     & 0.266395 - 0.089680 i  & 0.239947 - 0.281328  i &  0.204561 - 0.482084 i  \\
$0.08 $     & 0.262153 - 0.088252  i  & 0.236126 - 0.276847 i & 0.201304 - 0.474407 i   \\
\hline
\hline
$l=1$ \\
\hline
\hline
$\alpha$ &  $\omega_{3}$  & $\omega_{4}$  \\
\hline
$0.01$   &            0.171913 - 0.740918  i  & 0.109358 - 0.962654  i\\
$0.02$  &            0.170018 - 0.732753 i  & 0.108153 - 0.952046 i\\
$0.03$   & 0.168018 - 0.724132 i  & 0.106880 - 0.940845 i\\
$0.04 $    &        0.165911 - 0.715053 i  & 0.105540 - 0.929049 i\\
$0.05$  &         0.163698 - 0.705514 i  & 0.104132 - 0.916656 i \\
$0.06$   &         0.161379 - 0.695517 i  & 0.102657 - 0.903666  i\\
$0.07 $     &         0.158953 - 0.685062 i  & 0.101113 - 0.890082  i \\
$0.08 $     &         0.156421 - 0.674152 i & 0.099503 - 0.875907 i \\
\hline
\end{tabular}
\caption{Fundamental  quasinormal mode of the New EUP as a function of $\alpha$ calculated by the 3th order WKB approach($M=1$).}\label{tabl:em1}
\end{table*}

\begin{table*}
\begin{tabular}{|c|c|c|c|c||c||c||c||c||c|}
\hline

\hline
$l=1$\\
\hline
 $\alpha$ &$\omega_{0}$ & $\omega_{1}$ &  $\omega_{2}$  & \\
\hline
\hline
$0.01$   & 0.045077 - 0.015175 i& 0.040602 - 0.047604  i & 0.034614 - 0.081574 i \\
$0.02$   &0.040575 - 0.013660 i  & 0.036547 - 0.042849  i & 0.031157 - 0.073427 i \\
$0.03$   & 0.035095 - 0.011815 i  & 0.031611 - 0.037062 i & 0.026949 - 0.063510 i \\
$0.04 $   & 0.028828 - 0.009705 i  & 0.025966 - 0.030444 i & 0.022136 - 0.052168 i  \\
$0.05$  &0.022054 - 0.007424 i  & 0.019865 - 0.023290 i & 0.016935 - 0.039910 i  \\
$0.06$    & 0.015108 - 0.005086  i  & 0.013608 - 0.015955  i & 0.011601 - 0.027341  i \\
$0.07 $     & 0.008322 - 0.002802 i  &0.007496 - 0.008790 i & 0.006391 - 0.015061 i   \\
$0.08 $     & 0.001979 - 0.000666 i  &0.001782 - 0.002090 i &0.001520 - 0.003581 i   \\
\hline
\hline
$l=1$ \\
\hline
\hline
$\alpha$ &  $\omega_{3}$  & $\omega_{4}$  \\
\hline
$0.01$  &     0.026897 - 0.115929 i  & 0.017105 - 0.150611 i\\
$0.02$   &            0.024210 - 0.104342  i  & 0.015401 - 0.135569 i\\
$0.03$   & 0.020941 - 0.090251 i  & 0.013321 - 0.117260 i\\
$0.04 $  &        0.017201 - 0.074133 i  & 0.010942 - 0.096319 i\\
$0.05$   &         0.013159 - 0.056714 i  & 0.008371 - 0.073687 i\\
$0.06$   &         0.009015 - 0.038852 i  & 0.005735 - 0.050480 i\\
$0.07 $  &         0.004966 - 0.021402 i  & 0.003159 - 0.027807 i \\
$0.08 $   &         0.001181 - 0.005089 i  & 0.000751 - 0.006611 i \\
\hline
\end{tabular}
\caption{Fundamental quasinormal mode of the New EUP as a function of $\alpha$ calculated by the 3th order WKB approach ($M=6$).}\label{tabl:em2}
\end{table*}
The characteristic frequencies $\omega$ are complex. If the effective potential, given in equation \eqref{Veffpot}, initially makes a peak and then falls to a constant in the asymptotic region, one can compute the Q-N mode frequencies by employing the WKB approximation as done by Schutz et al. in \cite{kk6,kk7,kk8}. In this manuscript, we use the 3rd order WKB formula that is given in \cite{kk6}
\begin{equation}
\omega_{n}=\sqrt{\left(V_{0}+\Delta\right)-i\left(n+\frac{1}{2}\right) \sqrt{-2 V_{0}^{\prime \prime}}(1+\Omega)},
\end{equation}
where
\begin{equation}
\Delta=\frac{1}{8}\left(\frac{V_{0}^{(4)}}{V_{0}^{\prime \prime}}\right)\left(\frac{1}{4}+\gamma^{2}\right)-\frac{1}{288}\left(\frac{V_{0}^{\prime \prime \prime}}{V_{0}^{\prime \prime}}\right)^{2}\left(7+60 \gamma^{2}\right)
\end{equation}
and
\begin{equation}
\begin{aligned}
\Omega=&-\frac{1}{2 V_{0}^{\prime \prime}}\left\{\frac{5}{6912}\left(\frac{V_{0}^{\prime \prime \prime}}{V_{0}^{\prime \prime}}\right)^{4}\left(77+188 \gamma^{2}\right)\right.\\
&-\frac{1}{384}\left[\frac{\left(V_{0}^{\prime \prime \prime}\right)^{2}\left(V_{0}^{(4)}\right)}{\left(V_{0}^{\prime \prime}\right)^{3}}\right]\left(51+100 \gamma^{2}\right) \\
&+\frac{1}{2304}\left(\frac{V_{0}^{(4)}}{V_{0}^{\prime \prime}}\right)^{2}\left(65+68 \gamma^{2}\right) \\
&+\frac{1}{288}\left(\frac{V_{0}^{\prime \prime \prime} V_{0}^{(5)}}{\left(V_{0}^{\prime \prime}\right)^{2}}\right)\left(19+28 \gamma^{2}\right) \\
&\left.-\frac{1}{288}\left(\frac{V_{0}^{(6)}}{V_{0}^{\prime \prime}}\right)\left(5+4 \gamma^{2}\right)\right\}.
\end{aligned}
\end{equation}
We have $\gamma=n+1 / 2$ and $V_{0}^{(n)}$ means the $n$-order
derivative of the effective potential on the maximum point
$r_{*}$. Lately, the Wentzel-Kramers-Brillouin (WKB) approximation is generalized by Konoplya and Zhidenko to the sixth order \cite{kk9}. Note that the WKB approach is extended to higher orders in \cite{Konoplya:2003ii}, and there it is shown that it achieves a higher accuracy if the Padé approximant \cite{Matyjasek:2017psv,Konoplya:2019hlu,Churilova:2019jqx} is used.  However, the third-order WKB method has better accuracy when considering $n \leqslant l$ case in  small overtones number $n$.  Based on this, it can be seen from TABLE I that when the mass $(M=1)$, the real and imaginary parts of the Q-N frequency decrease with the increase of the EUP parameter. As we all know, the real part of the Q-N frequency represents the actual frequency of the black hole, while the imaginary part represents the attenuation of black hole oscillation. The imaginary part of the Q-N frequency is generally negative, indicating that the black hole oscillates steadily with the loss of time, indicating that the black hole is stable. For low-mass black holes, as the EUP parameter $\alpha$ increases, the decay of black holes becomes slower and slower. In TABLE II, for the massive black hole $(M=6)$, the real and imaginary parts of the Q-N frequency are significantly reduced, the EUP quantum correction has a certain influence on the stability of the massive Schwarzschild black hole. Similarly, the concept based on the minimum observable length, Anacleto etc. investigated the Q-N modes of the GUP corrected Schwarzschild black hole \cite{Anacleto}, the influence of GUP parameters $\beta$ on Q-N modes is similar to the EUP parameter $\alpha$ in our work. In quantum theory, coordinates and momentum are the two most basic operators, which indirectly indicate the correctness of our work.
\begin{figure*}
\resizebox{\linewidth}{!}{\includegraphics{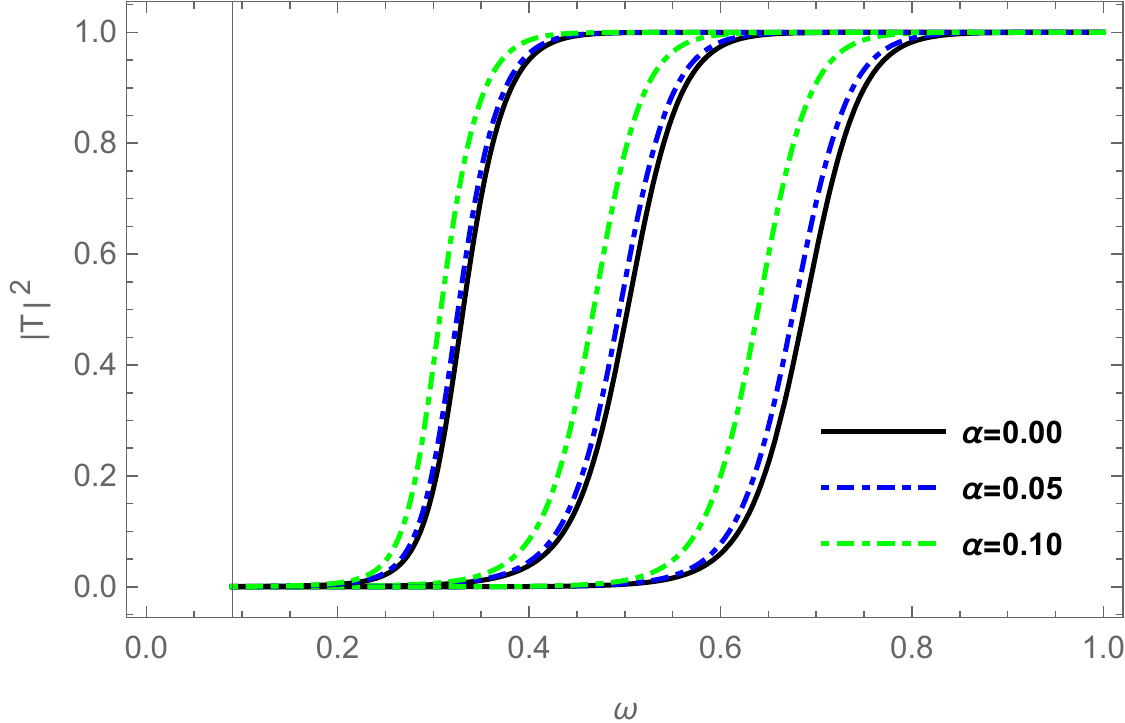},\includegraphics{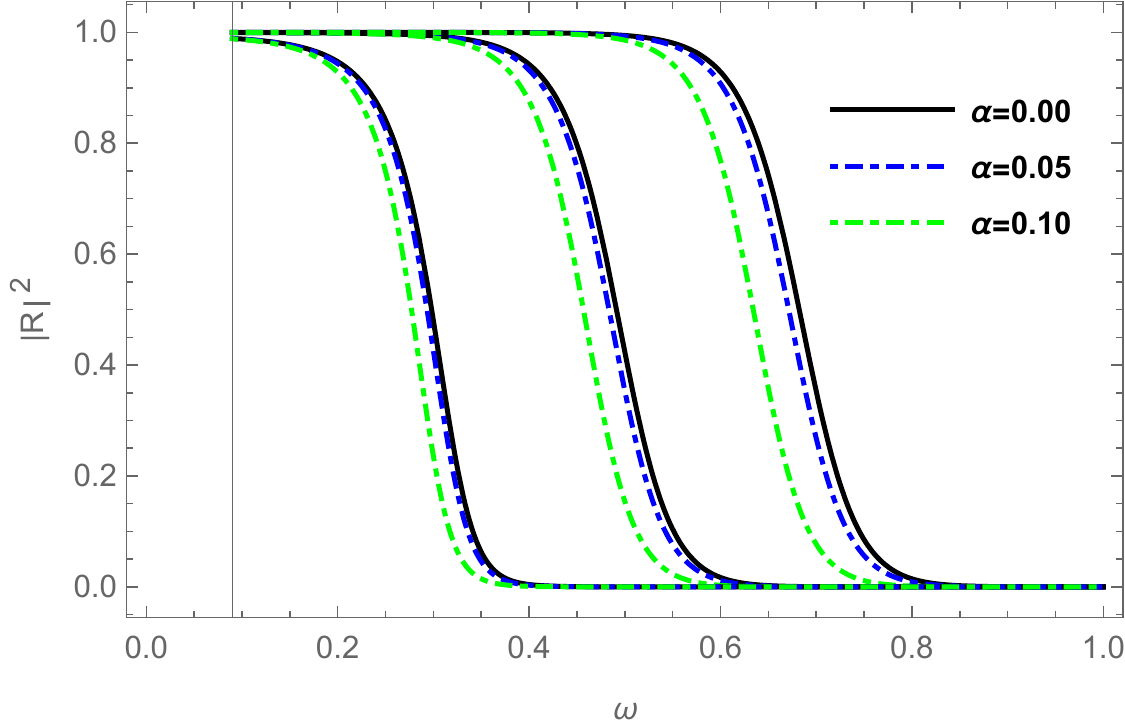}}
\caption{Transmission and reflection coefficients for three multipole l=1,2,3 (from left to right) and M=1. Left panel: Transmission coefficients. Right
panel: Reflection coefficients.}\label{fig:GreyBodyfromA1}
\end{figure*}
\begin{figure*}
\resizebox{\linewidth}{!}{\includegraphics{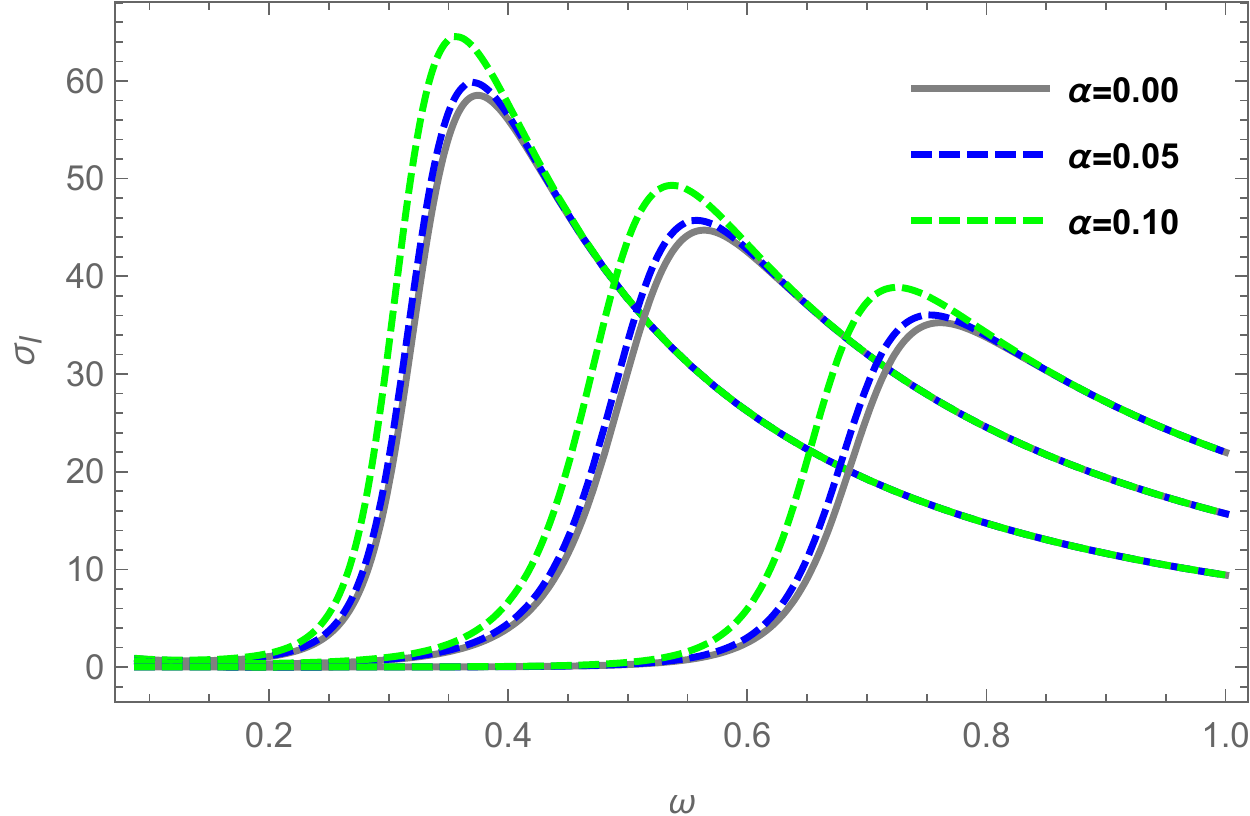},\includegraphics{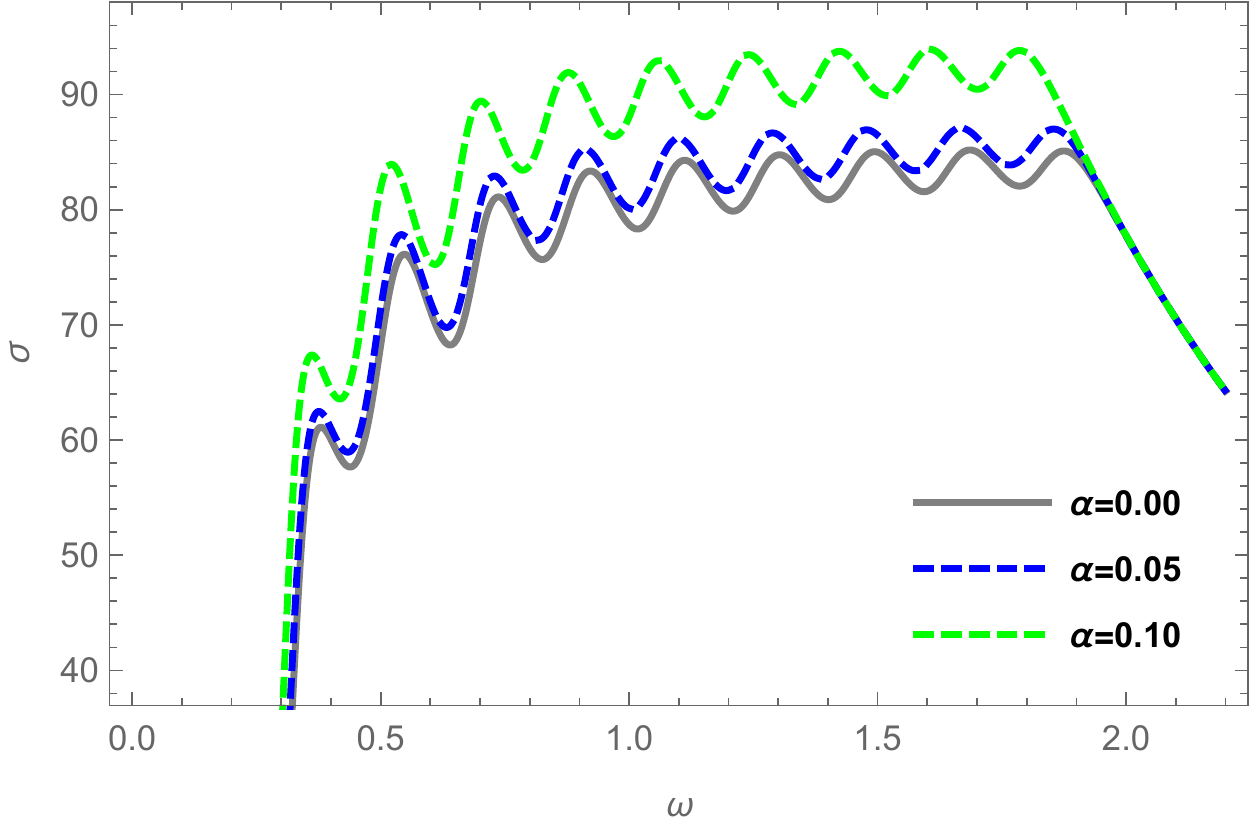}}
\caption{Absorption cross section for three multipole l=1,2,3 (from left to right) and M=1. left panel: Partial absorption cross section. Right
panel: All absorption cross section.}\label{fig:GreyBodyfromA2}
\end{figure*}
\section{Quantum effects of the scattering problem.}\label{sec:QNMs}
To estimate the portion of  the initial radiation near the event horizon, we will calculate the reflection and transmission coefficients, this is going to pave the way for our next work. We consider that the wave function satisfies the following boundary condition:
 \cite{kk6,kk7}
\begin{equation}\label{BC}
\begin{array}{ccll}
    \Psi &=& e^{-i\omega r_*} + R e^{i\omega r_*},& r_* \rightarrow +\infty, \\
    \Psi &=& T e^{-i\omega r_*},& r_* \rightarrow -\infty, \\
\end{array}%
\end{equation}
where $R$ and $T$ are the reflection and transmission coefficients. It will mean  waves to be incident from infinity.  It is well known that the effective potential exists in the form of a potential barrier that decreases monotonically at both points of infinity. The WKB method can be used to determine the reflection, $R$, and transmission, $T$, coefficients. $\omega^2$ is real so by considering the boundary conditions, given in equation \eqref{BC},  we can relate $K$ to the reflection and transmission coefficients as given in \cite{kk7}
\begin{equation}
|R|^{2}=\frac{\left|A_{\text {out }}\right|^{2}}{\left|A_{\text {int }}\right|^{2}}=\frac{1}{1+e^{-2 i \pi \mathcal{K}}}, \quad 0<|R|^{2}<1,
\end{equation}
and
\begin{equation}
|T_{l}|^{2}=\frac{\left|A_{t r}\right|^{2}}{\left|A_{i n t}\right|^{2}}=\frac{1}{1+e^{2 i \pi \mathcal{K}}}=1-|R|^{2}.
\end{equation}
Here, $K$ can be found from the following equation:
\begin{equation}
K-i \frac{\left(\omega^{2}-V_{0}\right)}{\sqrt{-2 V_{0}^{\prime \prime}}}-\sum_{i=2}^{i=6} \Lambda_{i}(K)=0.
\end{equation}
In FIG.4,  we plot the transmission and reflection coefficients versus the frequency. In the left panel, we consider three different multipole numbers namely $l=1$, $l=2$, and $l=3$ from left to right, and three EUP parameters as $\alpha=0$, $\alpha=0.05$, and $\alpha=0.1$. We observe that the increase of the EUP parameter affects the transmission and reflection coefficients.

Based on the transmission coefficient $T$, we define the partial absorption cross section, $\sigma_{l}$, and total absorption cross section, $\sigma$, respectively \cite{kk13}
\begin{equation}
\sigma_{l}=\frac{\pi(2 \ell+1)}{\omega^{2}}\left|T_{l}(\omega)\right|^{2},
\end{equation}
and
\begin{equation}
\sigma=\sum_{l} \frac{\pi(2 l+1)}{\omega^{2}}\left|T_{l}(\omega)\right|^{2}.
\end{equation}
In FIG.5, we plot the partial and all absorption for modes $l=1,2,3$. When the EUP parameter $\alpha$ is set to $0.05$, the absorption cross section is almost coincident with the Schwarzschild black hole ($\alpha=0$). The amplitude of the absorption cross section  increases with the increase of $\alpha$. Besides, we find that GUP and EUP parameters have a similar influence on the absorption cross section of the black hole \cite{kk31}.
\section{Conclusion}\label{sec:conclusions}
In this work, we employed the new EUP corrected Schwarzschild black
hole to investigate the Hawking temperature and evaporation rate. Based on the Stefan-Boltzmann law properties, the evaporation rate of the black hole is derived. To analyze the effect of the new EUP parameter $\alpha$ on the evaporation rate, we depicted the evaporation rate as the function of $M$ for the different EUP parameter values $(\alpha=0,0.05,0.1)$. We can observe that new EUP may drastically shorten the lifespan of the massive Schwarzschild black holes. To verify this conclusion, we continue to study the Q-N modes of the black hole. First, we derived the line element of the quantum corrected black hole and analyzed the influence of EUP on the time-
domain profiles for  the scalar field and the electromagnetic field disturbance,  we find as the EUP parameter increases, the oscillation and decay time of the black hole become slower. Then, we determined the Q-N modes numerically by the third-order WKB approximation. The result showed that the corrected-black hole does not lose its stability when the uncertainty parameter $\alpha$ is small, and the Q-N frequency decreases in the real and imaginary part with the increment of the EUP parameter. Furthermore, we studied the partial and all absorption cross section, and it shown that quantum effects significantly increase the total absorption cross-sectional area of Schwarzschild black holes.
\vspace{5mm}
\begin{acknowledgments}
We are very grateful to R. A. Konoplya and  S. Dey and S. Chakrabarti for kindly providing us the useful code and helpful correspondences.  Zheng-Wen Long is supported by the National Natural Science Foundation of China (Grant Nos. 11465006 and 11565009), and the Major Research Project of innovative Group of Guizhou province (2018-013). Bekir Can L\"{u}tf\"{u}o\u{g}lu is supported by the Internal  Project,  [2022/2218],  of  Excellent  Research  of  the  Faculty  of  Science  of Hradec Kr\'alov\'e University.
\end{acknowledgments}

\end{document}